\documentclass[twocolumn, floatfix, aps,showpacs]{revtex4}
\pdfoutput=1
\usepackage{graphicx}
\usepackage{amsmath}

\begin{document}
\title{Stark deceleration of CaF molecules in strong- and weak-field seeking states}
\author{T.E. Wall, J.F. Kanem, J. M. Dyne, J.J. Hudson, B.E. Sauer, E.A. Hinds and M.R. Tarbutt}
\address{Centre for Cold Matter, Blackett Laboratory, Imperial College London, Prince Consort Road, London, SW7 2AZ, UK.}

\begin{abstract}

We report the Stark deceleration of CaF molecules in the strong-field seeking ground state and in a weak-field seeking component of a rotationally-excited state. We use two types of decelerator, a conventional Stark decelerator for the weak-field seekers, and an alternating gradient decelerator for the strong-field seekers, and we compare their relative merits. We also consider the application of laser cooling to increase the phase-space density of decelerated molecules.

\end{abstract}

\maketitle

\section{Introduction}
A Stark decelerator is a device for slowing down pulses of cold molecules \cite{firstStarkMeijer}. The force that acts on the molecules is the spatial gradient of their Stark shift arising from the inhomogeneous electric field of the decelerator. The ability of a Stark decelerator to deliver cold polar molecules with a centre-of-mass velocity at or near zero has proven exceedingly useful for a wide variety of experiments including high precision spectroscopy \cite{vanVeldhoven(1)04}, measurements of the lifetimes of long-lived molecular states \cite{Meerakker(1)05, Gilijamse(1)07}, low temperature collision studies \cite{Gilijamse(1)06, Scharfenberg(1)11}, and tests of fundamental physics \cite{Hudson(1)06, Bethlem(1)08}.

To date, only relatively light polar molecules in weak-field seeking states have been decelerated to low speed by the Stark deceleration method \cite{firstStarkMeijer, Bethlem(1)02, Meerakker(2)05, Hudson(2)06, Hoekstra(1)07, Bucicov(1)08, Tokunaga(1)09}. For many applications it would be useful to extend the range of slow molecules available by applying the method to heavier molecules. This is particularly relevant for molecule-based tests of fundamental physics, such as the measurement of the electron's electric dipole moment \cite{Hudson(1)02} and tests of parity violation in nuclei \cite{DeMille(1)08} and chiral molecules \cite{Darquie(1)10}, all of which require heavy molecules to reach high precision and could benefit from the use of slower molecules \cite{Tarbutt(1)09}.

The kinetic energy to be removed by the decelerator is proportional to the molecular mass, so heavier molecules require more deceleration stages. Fortunately, the decelerator focusses the bunches of molecules both longitudinally \cite{Bethlem(1)00} and transversally \cite{Meerakker(1)06}, so the bunches do not spread out as more stages are added. This important focussing property applies only to molecules in weak-field seeking states, since they are naturally attracted to the axis of the machine where the field is lowest. Focussing of strong-field seekers in all three directions is not possible because of the impossibility of creating an electric field maximum in free space \cite{Auerbach(1)66}. Unfortunately, the low-lying states of heavy molecules are all strong-field seeking in the large electric fields needed for a decelerator. This is a major obstacle to extending Stark deceleration to heavier molecules. There are two possible ways to circumvent this difficulty. One option is to prepare the molecules in a higher-lying rotational state which does have a weak-field seeking component at the relevant fields, and then decelerate them in the usual way. The heavier the molecule the greater the rotational excitation will need to be in order to find a suitable weak-field seeking state. The second option is to an use an alternating gradient (AG) decelerator \cite{Bethlem(2)02, Tarbutt(1)04, Bethlem(1)06} which works with molecules in strong-field seeking states. In this type of decelerator each deceleration stage focusses the molecules in one transverse direction and defocusses them in the other, the focussing and defocussing directions alternating from one stage to the next. This series of focussing and defocussing lenses can, when properly arranged, provide net focussing in both transverse directions. An AG decelerator can be used for molecules in any state, including the ground state which always has the largest Stark shift, and its application to heavy molecules has been demonstrated \cite{Tarbutt(1)04, Wohlfart(1)08}.

In this paper we describe the deceleration of CaF molecules in both weak-field seeking and strong-field seeking states. For the former case, we use CaF molecules in a weak-field seeking component of the fourth rotationally excited state ($N=4$) so as to maximize the Stark shift, and we use a decelerator that follows a conventional design - we call it a WF-decelerator. For decelerating strong-field seeking CaF we use molecules in the ground rotational state and an AG decelerator. We compare these two strategies for decelerating heavy molecules. We also consider the prospects of laser cooling CaF so as to increase the number and phase-space density of decelerated molecules.

\section{Experimental set-up}

\begin{figure}[tb]
\centering
\includegraphics[width=7.5cm]{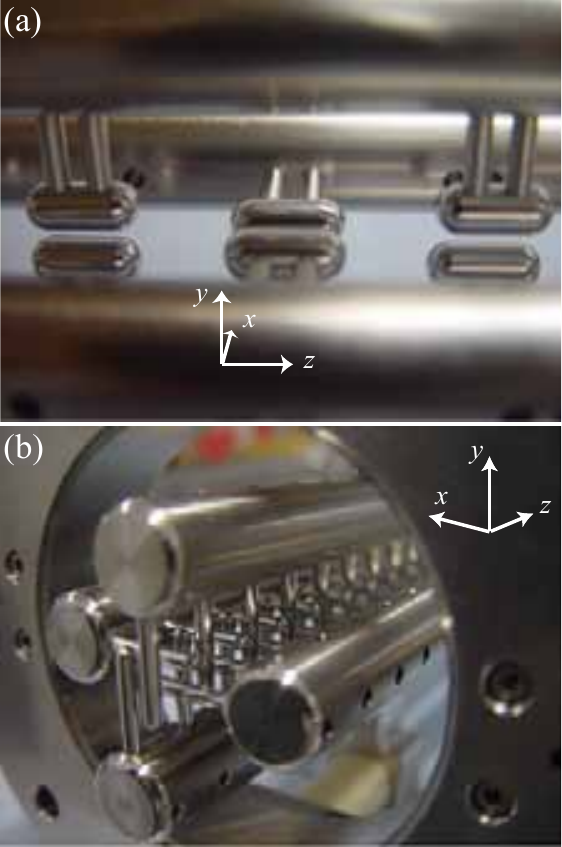}
\caption{Photographs of the two decelerators. (a) Three stages of the AG decelerator. (b) The first few stages of the WF decelerator.}
\label{AGDecelPhoto}
\end{figure}

A pulsed supersonic beam of CaF is produced by laser ablation of Ca into a pulsed supersonic carrier gas (Ar, Kr or Xe) containing a small fraction of SF$_6$ \cite{Tarbutt(1)02}. The translational and rotational temperatures of the molecules are both approximately 3\,K. Their speed is usually a little higher than the terminal supersonic speed of the carrier gas. The molecules pass through a skimmer, then through the decelerator, and are finally detected 805\,mm from the source by time-resolved cw laser-induced fluorescence on a chosen rotational component of the $A^{2}\Pi_{1/2} - X^{2}\Sigma^{+} (0 - 0)$ transition at 606\,nm. The experiment runs at a repetition rate of 10\,Hz. Further details of the methods we use to produce and detect cold CaF molecules are given in \cite{Wall(1)08}.

The AG decelerator, shown in Fig.\,\ref{AGDecelPhoto}(a), consists of 21 deceleration stages, each of which is also a lens that focusses in one transverse direction and defocusses in the other. Each of these lenses is formed by a pair of stainless steel cylindrical electrodes, 14\,mm long and 6\,mm in diameter, with hemispherically-rounded ends of radius 3\,mm, their axes parallel to the $z$-axis, and their surfaces separated by 2\,mm. The centre-to-centre spacing of the lenses (along $z$) is 24\,mm, and successive lenses are rotated through $90^{\circ}$ about the $z$-axis. When the electrode pair lies in the $x z$-plane the lens focusses strong-field seeking molecules along $y$ and defocusses them along $x$. The following lens has its electrodes in the $y z$-plane and does the opposite. Each electrode is attached by a pair of dowels to one of four 16\,mm-diameter stainless steel support rods, 504\,mm in length. Each support rod is attached by insulating macor stand-offs to two stainless steel support rings, one at either end of the decelerator.

The WF decelerator, shown in Fig.\,\ref{AGDecelPhoto}(b), consists of 100 deceleration stages. Each stage is formed by two parallel stainless steel cylindrical electrodes of diameter 3\,mm, with their axes parallel to either $x$ or $y$, and their surfaces separated by 2\,mm. The centre-to-centre spacing of the stages is 6\,mm along $z$, and successive stages are rotated through $90^{\circ}$ about the $z$-axis. When the electrodes have their axes along $x$ they focus weak-field seeking molecules along $y$, and do nothing along $x$. For the electrodes aligned along $y$ the opposite is true. Each electrode is pushed into one of four 16\,mm-diameter stainless steel support rods, 594\,mm in length, and these support rods are attached by insulating alumina stand-offs to two stainless steel support rings.

For both decelerators, four independent 20\,kV switches are used to switch the high voltages applied to each of the support rods between two values, $\pm V_{\text{HI}}$ and $\pm V_{\text{LO}}$, with rise and fall times of approximately 500\,ns. The support rings are grounded. For the AG decelerator, $\pm V_{\text{LO}}$ is zero. To avoid nonadiabatic transitions in the WF decelerator \cite{Wall(1)10}, $V_{\text{LO}}$ is not zero but is still far smaller than $V_{\text{HI}}$. After high voltage conditioning, both decelerators are able to support $\pm 20$\,kV across the 2\,mm gaps between electrode surfaces.

\section{Properties of the two decelerators}

\begin{figure}[tb]
\centering
\includegraphics[width=7.5cm]{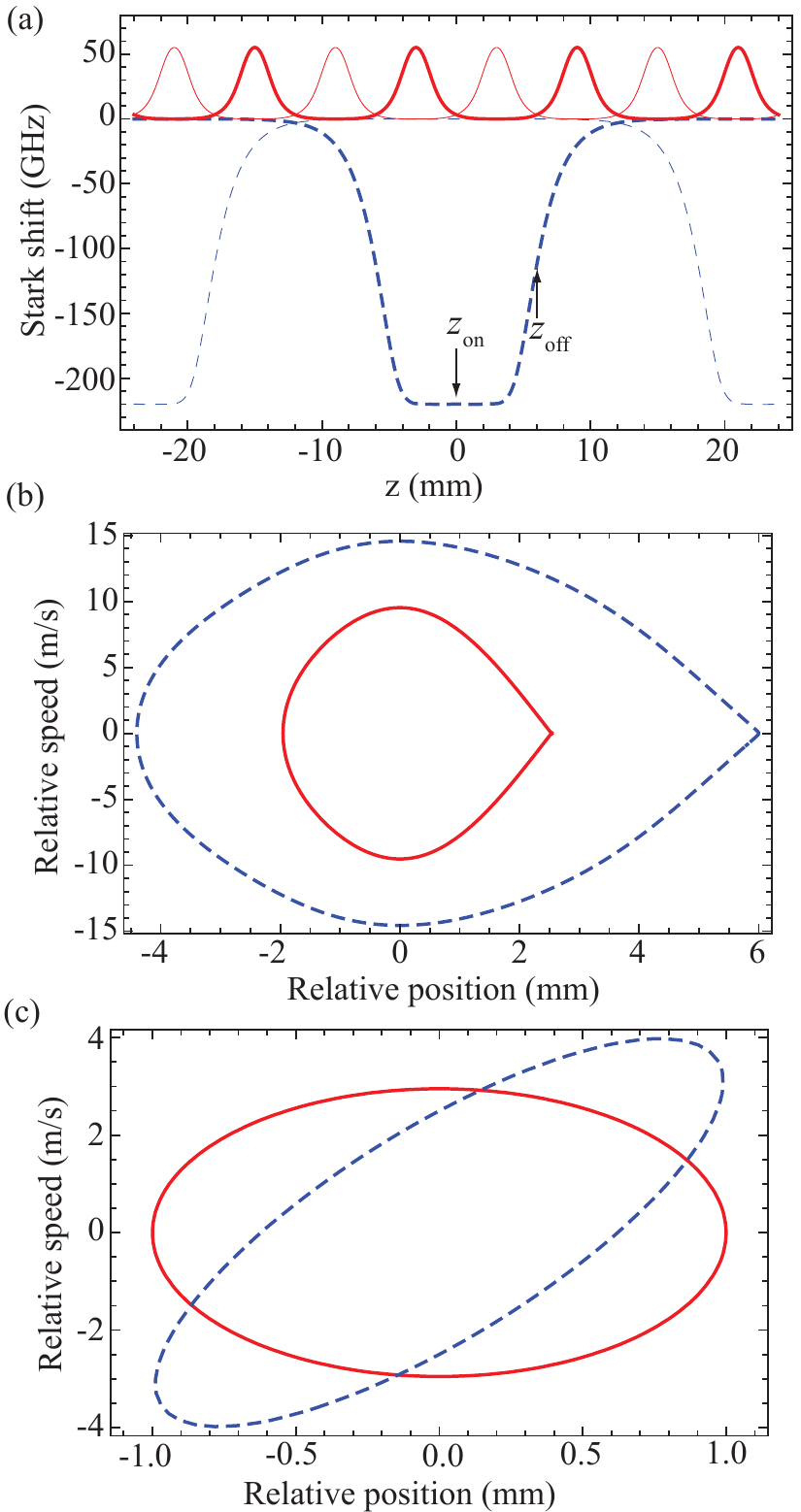}
\caption{Key information for deceleration of CaF in the two decelerators: WF (solid red lines) and AG (dashed blue lines). Both decelerators are charged to $\pm 20$\,kV. The molecules are in the $(N,M)=(4,0)$ rotational state for the WF decelerator and in the $(0,0)$ state for the AG decelerator. (a) Stark shift versus longitudinal position with even stages charged and odd stages grounded (bold) and vice versa (thin). (b) Longitudinal acceptance when the energy loss per stage is half the maximum possible value. (c) Transverse acceptance when the energy loss per stage is half the maximum possible value. For the AG decelerator we show the acceptance area at the entrance of a focussing lens, we took an effective lens length of 5.5\,mm, an effective drift length of 18.5\,mm and a forward speed of 430\,m/s.}
\label{AGvsWF}
\end{figure}

As is usual, we introduce a reduced position $\theta = \pi z/L$ (modulo $2\pi$) where $L$ is the distance between deceleration stages. The decelerators can be in one of four high voltage configurations: (i) even stages at $\pm V_{\text{HI}}$, odd stages at $\pm V_{\text{LO}}$, (ii) odd stages at $\pm V_{\text{HI}}$, even stages at $\pm V_{\text{LO}}$, (iii) all stages at $\pm V_{\text{LO}}$, and (iv) all stages at $\pm V_{\text{HI}}$. The WF decelerator is switched back and forth between configurations (i) and (ii) in a sequence generated such that a molecule with a chosen initial speed $v_{i}$ reaches the same point relative to the periodic array every time the decelerator is switched. The value of $\theta$ at this point is termed the synchronous phase angle, $\phi_{0}$ \cite{Bethlem(1)00}. The switching sequence for the AG decelerator is slightly more complicated - the decelerator is switched into state (i) when the synchronous molecule reaches the position $z_{\text{on}}$ inside an even stage, to state (iii) when it reaches the position $z_{\text{off}}$, to state (ii) when at $z_{\text{on}}$ in an odd stage, and back to state (iii) when it reaches $z_{\text{off}}$ again.

In Fig.\,\ref{AGvsWF} we compare the basic properties of the two decelerators. Here, we take the electrodes to be charged to $\pm 20$\,kV so that the field on the beamline is approximately 200\,kV/cm, and we take the CaF molecules to be in the $(N,M)=(0,0)$ state for the AG decelerator and in the $(4,0)$ state for the WF decelerator. Figure \ref{AGvsWF}(a) shows the Stark potential as a function of position ($z$) along the beamline for configurations (i) and (ii). Although the electric field on the beamline is approximately the same in the two decelerators, the depth of the Stark potential is about 4 times higher in the AG than in the WF because of the larger Stark shift of the ground state. However, the stages are packed 4 times closer together in the WF and so the deceleration per unit length of decelerator is approximately the same in the two cases. Figure \ref{AGvsWF}(b) shows the longitudinal acceptance areas of the two decelerators, calculated analytically in both cases using the relevant effective potentials \cite{Bethlem(1)00, Tarbutt(2)09}. In both cases, the energy loss per stage is chosen to be half the maximum possible value. The longitudinal acceptance of the AG is larger than the WF because the stages are longer and the the Stark shift is larger. Figure \ref{AGvsWF}(c) shows the acceptance areas of the two decelerators in one of the two transverse directions. For the WF decelerator this is obtained by calculating the mean transverse spring constant \cite{Meerakker(1)06}, while for the AG decelerator it is found by using the Courant-Snyder formalism for AG focussing \cite{Bethlem(1)06, nonlinDynamics, Tarbutt(2)09}. We find that the transverse acceptance areas are approximately the same for the two decelerators. In the AG decelerator each lens focusses/defocusses strongly, because of the large Stark shift. The net effect of the alternating lenses is to focus the molecules, and this can be described as an effective focussing lens whose power is considerably smaller than that of the individual lenses \cite{nonlinDynamics}. In the WF decelerator the molecules are focussed throughout, but the focussing is relatively weak because of the smaller Stark shift and because, whenever the molecules approach a region of strong field where the focussing is strongest the decelerator is switched to make the field small again.

We have used approximate methods to calculate the phase-space acceptance areas shown in Fig.\,\ref{AGvsWF}. In particular we have used an effective potential to describe the longitudinal motion of molecules relative to that of the synchronous molecule, we have assumed that the transverse forces are harmonic, and that the longitudinal and transverse motions are uncoupled. This last approximation is particularly poor, the motions in the longitudinal and transverse directions being quite strongly coupled in both the WF decelerator \cite{Meerakker(1)06} and the AG decelerator \cite{Bethlem(1)06}. The true acceptances will be smaller as a consequence, but still of the same order of magnitude as obtained from these approximate methods (see \cite{Tarbutt(1)09} for example.)

\section{AG deceleration results}

\begin{figure}[tb]
\centering
\includegraphics[width=8cm]{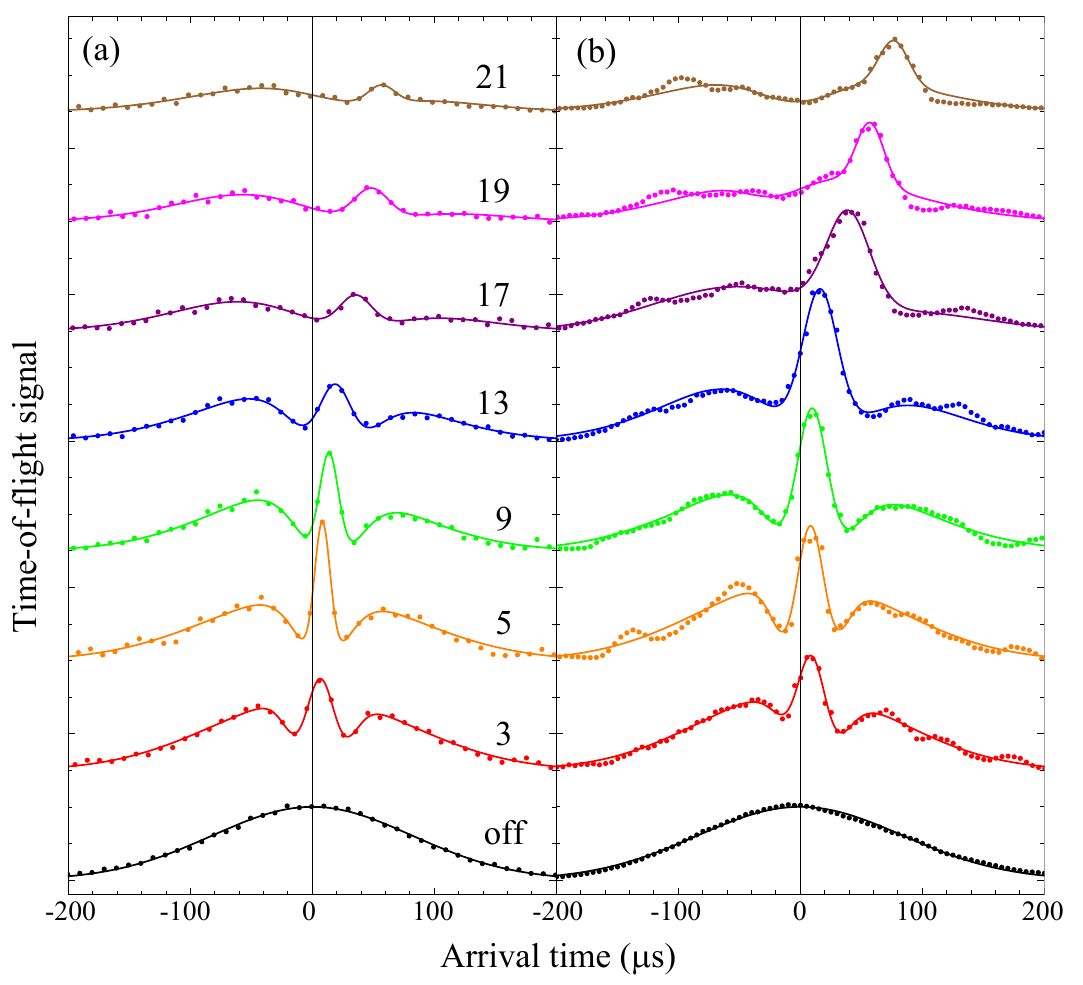}
\caption{(a) Points: Measured time-of-flight profiles of ground-state CaF with the last $n$ stages of the AG decelerator being used. The parameters are $V_{\text{HI}} = 20$\,kV, $V_{\text{LO}} = 0$, $z_{\text{on}}=0$\,mm, $z_{\text{off}}=6$\,mm, $v_{i}=433$\,m/s. The bottom profile was taken with the decelerator turned off. Lines: Triple Gaussian fits to the data. (b) Points: Simulated time-of-flight profiles to match the experiments. Lines: Triple Gaussian fits.}
\label{all2430AGDecelData}
\end{figure}

In this section we discuss the deceleration of ground state CaF from an initial speed of $v_i=433$\,m/s. The AG decelerator is used with $V_{\text{HI}}= 20$\,kV, $V_{\text{LO}}= 0$, $z_{\text{on}}=0$\,mm and $z_{\text{off}}=6$\,mm, as indicated in Fig.\,\ref{AGvsWF}(a). The deceleration switching sequence is applied on every even-numbered pulse of the experiment, while on every odd-numbered pulse the decelerator is turned off. Figure \ref{all2430AGDecelData}(a) shows how the time-of-flight profile of the molecules changes as the number of deceleration stages used in the experiment, $n$, is increased. It is the {\it last} $n$ stages that are used, the decelerator being off until the synchronous molecule is $n$ stages from the decelerator exit. To eliminate the effects of drifts in the source flux, each profile has been normalized to the amplitude of the corresponding decelerator-off profile obtained (almost) simultaneously. The zero of time is the arrival time of the synchronous molecule (with speed $v_i$) when the decelerator is off. After a few stages the phase-stable molecules have been bunched about the synchronous molecule, resulting in a narrow peak in the measured time-of-flight profiles. As $n$ is increased the narrow peak moves to later arrival times, showing that the speed of this bunch has been reduced.

\begin{figure}[tb]
\centering
\includegraphics[width=7.5cm]{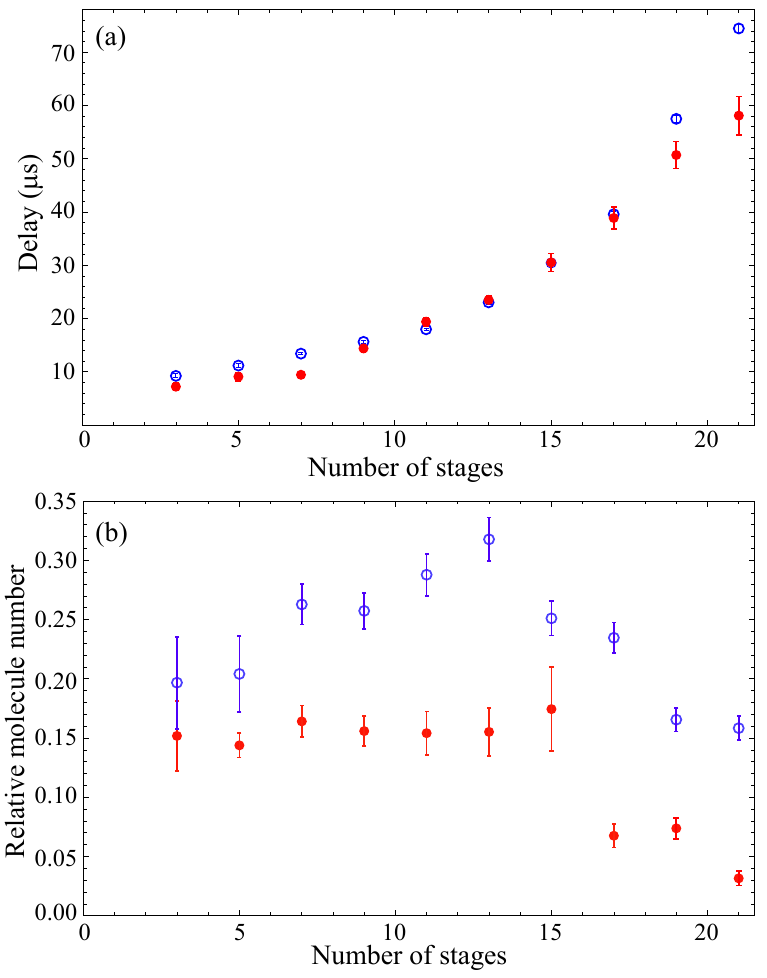}
\caption{Comparison of experimental (filled circles, red) and simulated (open circles, blue) results for the decelerated bunch, as determined by the parameters of a three-Gaussian fit. The error bars are the uncertainties on the fitted parameters; when not visible, they are smaller than the size of the points. (a) Delay in arrival time as a function of $n$. (b) Number of molecules in the decelerated bunch as a function of $n$, normalized to the total number of molecules measured when the decelerator is off.}
\label{AGResultsSummary}
\end{figure}

Figure \ref{all2430AGDecelData}(b) shows the time-of-flight profiles obtained by simulating the motion of molecules through the decelerator, using a field map obtained from a finite element model and removing any molecule that crashes into an electrode. The simulation results agree fairly well with the experimental data. To make a quantitative comparison, we fit each of the experimental and simulated datasets to a model which is a sum of three Gaussians. In this model, one Gaussian ($G_1$) represents the broad background of undecelerated molecules, a second ($G_2$) represents the decelerated bunch, and a third ($G_3$), with negative amplitude, represents the hole in the undecelerated distribution where the decelerated bunch would otherwise have been. These fits are shown as lines in figure \ref{all2430AGDecelData}. We see that this model fits well to the data, except for some modulations in the undecelerated background that the model does not capture. These modulations, which are due to bunching of molecules in the neighbouring deceleration stages, are most clearly seen in the simulated data for small $n$, but are also present to a lesser extent in the experimental data, and in the data for higher $n$. As one would expect, the central arrival times of $G_1$ and $G_3$ have no significant dependence on $n$, while the arrival time of $G_2$ increases with $n$. This is shown in Fig.\,\ref{AGResultsSummary}(a) where we plot, as a function of $n$, the delay in arrival time between the decelerated bunch and the arrival time of the synchronous molecule when the decelerator is off. Note that for the relatively small reductions in speed obtained here, the narrow bunch of molecules represented by $G_2$ is actually a mixture of phase-stable molecules and some molecules that are not phase-stable but have not yet separated from the phase-stable bunch. The experimental delays are similar to the predictions of the simulations but deviate for $n=19$ and 21 where the simulations predict a larger delay than is measured. We do not know the reason for this discrepancy. In the experiments for $n=21$ the phase stable bunch is decelerated from 433\,m/s to 399\,m/s, corresponding to a removal of 2.1\,THz of kinetic energy, or 15\% of the initial kinetic energy.

Figure \ref{AGResultsSummary}(b) shows how the area of the decelerated bunch (i.e.~of $G_{2}$) depends on $n$ for both the experimental and simulated data. We have normalized these areas to the total area of the signal obtained when the decelerator is off. The simulations predict an increase in the signal ratio up to $n=13$ and then a decrease at higher $n$. When $n$ is small, the distance from the source to the effective entrance of the decelerator is large, and the range of transverse speeds that can enter the decelerator is smaller than the range the decelerator can accept. The decelerator's acceptance is thus not filled. As $n$ increases the effective entrance moves closer to the source, the acceptance is more completely filled and so the number of molecules in the decelerated bunch increases. At around $n=13$ the acceptance is completely filled and further increasing $n$ now reduces the signal ratio because there is a loss mechanism at work in the decelerator, as we discuss below. The measured signal ratio for the decelerated bunch is always smaller than the simulations predict, remaining constant up to $n=15$, and then dropping. This is the behaviour we would expect if the continuous losses within the decelerator are larger than expected. We note that there is a similar reduction with increasing $n$ in the transmission of the undecelerated molecules.

\begin{figure}[tb]
\centering
\includegraphics[width=7cm]{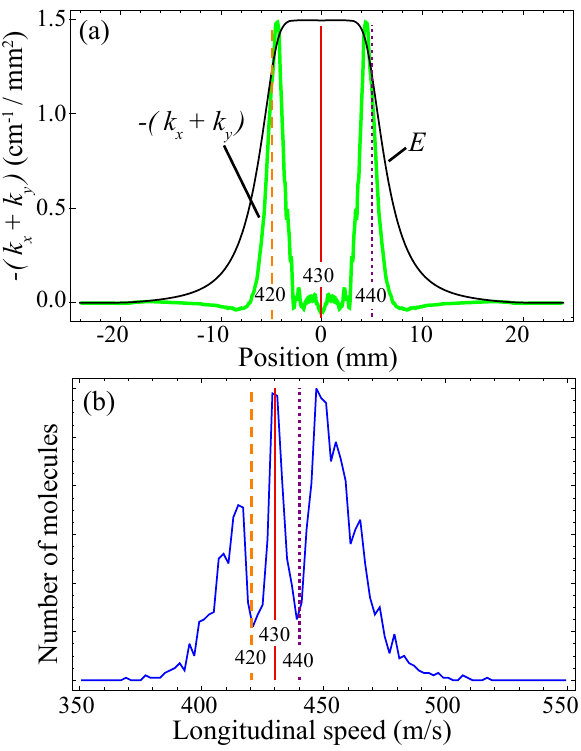}
\caption{(a) The quantity $-(k_{x}+k_{y})$ plotted as a function of distance $z$ along the AG decelerator, indicating the regions of the lens where the defocussing is far stronger than the focussing. For reference, the electric field as a function of $z$ is also plotted (on a different scale). (b) Simulation result showing the distribution of initial velocities that are successfully transmitted through the decelerator. Molecules with three velocities are indicated by vertical lines and are discussed in the text. Their central positions inside the first lens, during the time that it is turned on, are shown by the equivalent vertical lines in (a).}
\label{lostFigure}
\end{figure}

The transverse forces at the ends of each lens are responsible for these continuous losses. In an ideal AG decelerator, where the transverse forces are linear in the off-axis displacement with equal and opposite force constants, the transverse acceptance does not change as more lenses are added, at least not once the decelerator length is longer than the wavelength of the macromotion. As has been discussed previously \cite{Bethlem(1)06}, the ends of the AG decelerator electrodes are detrimental to proper AG focussing because the magnitude of the defocussing force constant $k_{y}=-\partial F_{y}/\partial y$ greatly exceeds the magnitude of the focussing force constant $k_{x}=-\partial F_{x}/\partial x$ near these ends, $F_{x,y}$ being the components of the force in the focussing ($x$) and defocussing ($y$) directions. This is shown in Fig.\,\ref{lostFigure}(a), where we plot the quantity $-(k_{x}+k_{y})$ as a function of position along the axis, $z$. For reference, the electric field along $z$ is shown on the same plot. While the focussing and defocussing force constants are well-balanced inside the lens, the defocussing is far stronger than the focussing in the region near the ends of the lens. Figure \ref{lostFigure}(b) demonstrates the impact of these end effects. Here, we have simulated AG deceleration with $n=17$,  $V_{\text{HI}} = 20$\,kV, $V_{\text{LO}} = 0$, $z_{\text{on}}=0$\,mm, $z_{\text{off}}=6$\,mm and $v_{i}=436$\,m/s. The speed distribution entering the decelerator is a Gaussian centred at 436\,m/s. From the simulation we can select out those molecules that are transmitted by the decelerator and then look at their distribution of initial speeds. It is this distribution that is plotted in Fig.\,\ref{lostFigure}(b). We see that the transmission depends very strongly on the speed. We draw attention to molecules with three specific velocities, 420\,m/s, 430\,m/s and 440\,m/s, indicated by the vertical lines, where the transmission is bad, good and bad, respectively. Each of these molecules passes through a 6\,mm-wide region of the first lens during the time when it is switched on. For each molecule, the centre of this region is indicated by a vertical line in Fig.\,\ref{lostFigure}(a). The molecules that are poorly transmitted are the ones that see the two bad regions of this first lens. The simulations show that these molecules repeatedly experience the bad regions as they move from one lens to the next, and that is why they tend to be lost. The 430\,m/s molecule sees only the good part of the first lens, and to a large extent avoids the bad regions of the other lenses too, and so is transmitted well by the decelerator. Unfortunately, the bad region at the exit of each lens is also the region where the molecules are decelerated, and so the decelerated molecules necessarily have to move through this bad region and tend to be thrown out of the decelerator for this reason.

Our simulations use an accurate map of the field, calculated numerically using a finite element model, but the experimental data shows more molecule loss than in the simulations. We do not know the reason for this. Other AG deceleration and focussing experiments also report greater losses than expected, the suspected cause being electrode misalignments \cite{Bethlem(2)02, Wohlfart(1)08, Filsinger(1)08, Wall(1)09}.

\section{WF deceleration results}

In this section we discuss the WF deceleration of CaF molecules that emerge from the source in the $(N,M)=(4,0)$ state and with an initial speed of 340\,m/s. For these experiments, it is important to suppress nonadiabatic transitions to other $(4,M)$ states which may be driven by the rotation of the electric field when the decelerator switches, especially in regions where the electric field is small \cite{Wall(1)10}. These unwanted transitions can be suppressed by ensuring that the Stark splitting between the various $M$ states is much larger than the angular frequency at which the field rotates. For this reason, instead of switching between high voltage and ground, we switch between $V_{\text{HI}}$ and a bias voltage of $V_{\text{LO}}\simeq 1$\,kV. This size of bias voltage eliminates losses due to nonadiabatic transitions and makes no significant difference to the energy loss per stage.

\begin{figure}[tb]
\centering
\includegraphics[width=8.2cm]{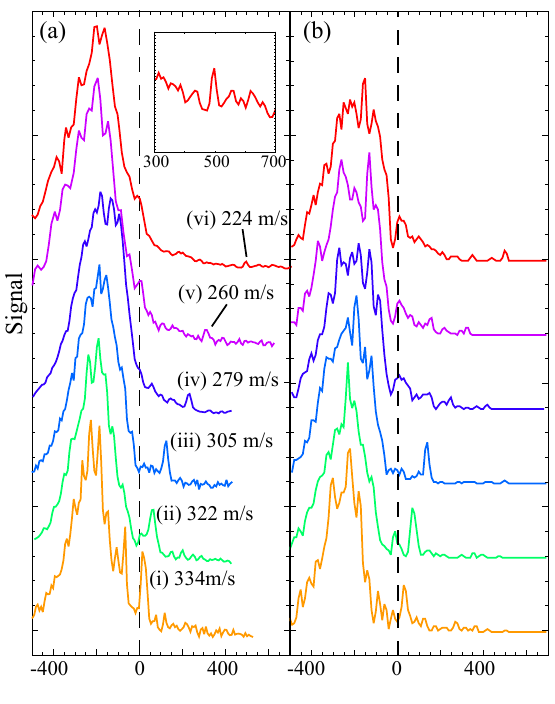}
\caption{(a) Measured time-of-flight profiles of CaF in the $N=4$ rotational state following deceleration in the WF decelerator. The zero of time is the arrival time of the synchronous molecule when the decelerator is off. For all the data, $v_{i} = 340$\,m/s, $V_{\text{LO}}=1.0$\,kV. The high voltage ($V_{\text{HI}}$) and synchronous phase angles ($\phi_0$) for the individual traces are (i) 18\,kV, $15^\circ$, (ii) 18\,kV, $30^\circ$, (iii) 18\,kV, $45^\circ$, (iv) 18\,kV, $60^\circ$, (v) 20\,kV, $60^\circ$, (vi) 20\,kV, $75^\circ$. The inset shows the decelerated bunch for trace (vi) in more detail. (b) Numerical simulations of these same experiments but with $V_{\text{LO}}=0$ and non-adiabatic transitions neglected.}
\label{allWFDecelData}
\end{figure}

Figure \ref{allWFDecelData} shows the results of these experiments, along with the associated numerical simulations. The laser-induced fluorescence signal from molecules in $N=4$ is plotted versus arrival time for several different values of the synchronous phase angle and the high voltage $V_{\text{HI}}$. The zero of time is the arrival time of the synchronous molecule when the decelerator is turned off, or when it is operated at $\phi_{0}=0$. As $\phi_{0}$ increases the decelerated bunch moves to later arrival times, corresponding to slower speeds, but also reduces in amplitude because the longitudinal phase space acceptance of the decelerator decreases as $\phi_{0}$ increases. In the uppermost trace, we have set $\phi_0=75^\circ$ and $V_{\text{HI}}=20$\,kV. In this case the molecules are decelerated to a final speed of 224\,m/s, corresponding to an energy reduction of 4.9\,THz or a removal of 57\% of the initial kinetic energy. The simulated time-of flight profiles shown in Fig.\,\ref{allWFDecelData}(b) agree well with the experimental data. In particular, the measured arrival time and size of the decelerated bunch are the same as predicted by the simulations.

\begin{figure}[tb]
\centering
\includegraphics[width=7cm]{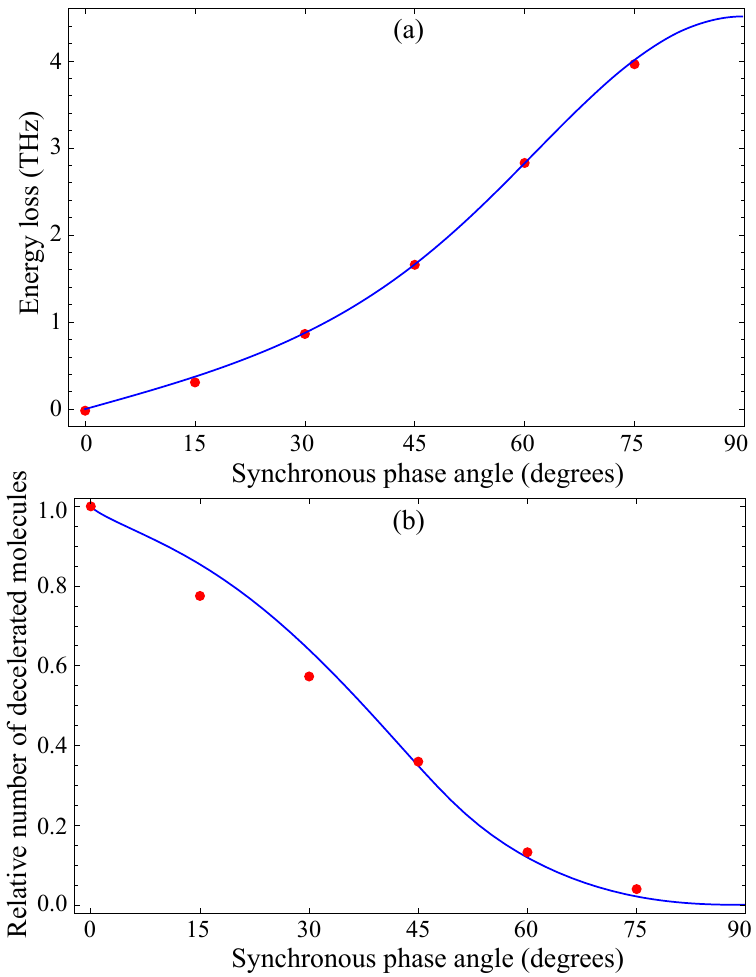}
\caption{Results from the WF decelerator operated at $V_{\text{HI}}=18$\,kV. (a) Points: Measured kinetic energy loss for the decelerated bunch versus synchronous phase angle. Line: Expected energy loss calculated from the Stark potential. (b) Points: Measured number of decelerated molecules versus synchronous phase angle, normalized to the number at $\phi_{0}=0$. Line: Longitudinal acceptance versus synchronous phase angle.}
\label{WFResultsSummary}
\end{figure}

Figure \ref{WFResultsSummary} (a) shows how the measured energy loss changes with $\phi_{0}$ when $V_{\text{HI}}=18$\,kV. Here, we have determined the energy loss from the measured arrival time of the decelerated bunch and the measured distances from source to decelerator and decelerator to detector, by assuming that the decelerator applies a constant force to the molecules over its entire length. Since there are many deceleration stages, each applying the same net deceleration, this is a good approximation. The energy loss follows the prediction obtained from the Stark potential plotted in Fig.\,\ref{AGvsWF}(a). In Fig.\,\ref{WFResultsSummary}(b) we show the number of decelerated molecules as a function of $\phi_{0}$, normalized to the number at $\phi_{0}=0$ and compare this to the (similarly normalized) longitudinal acceptance. As $\phi_{0}$ increases the number of decelerated molecules decreases because the longitudinal acceptance decreases. While the experimental results follow the trend of the longitudinal acceptance, there is some deviation from this simple model. This is because the transverse acceptance also has some dependence on $\phi_{0}$, through the coupling of the transverse and longitudinal motions. This coupling tends to reduce the number of decelerated molecules at low phase angles and to increase it at high phase angles \cite{Meerakker(1)06}, as we indeed observe in Fig.\,\ref{WFResultsSummary}(b).

\section{Prospects for combining Stark deceleration and laser cooling}

Laser cooling of a diatomic molecule has recently been demonstrated for the first time \cite{Shuman(1)10}. Laser cooling is feasible if the molecule has a short-lived state that can be excited with a convenient laser and that decays (exclusively) to the ground state with a Franck-Condon factor close to unity. Then, only a few excitation lasers are needed to address all relevant vibrational components of the electronic transition \cite{Shuman(1)09}. The $\text{A}^{2}\Pi_{1/2}-\text{X}^{2}\Sigma^{+} (0-0)$ transition of CaF has all the required properties for laser cooling \cite{Wall(1)08}. It may be possible to both cool and decelerate the CaF beam using a frequency-chirped counter-propagating laser. To decelerate from 340\,m/s requires the molecule to scatter about 30000 photons, and it is not yet clear how many of the vibrational transitions would need to be addressed to achieve this since the Franck-Condon factors are not known well enough. An alternative approach is to combine laser cooling and Stark deceleration. Laser cooling prior to deceleration increases the phase space density of the molecular pulse and so increases the number of molecules that can be captured by the decelerator. The decelerator is then used to bring the molecules to rest. Since the cooling only needs to change the molecule speeds by about 10\,m/s to compress the velocity distribution substantially, the required number of scattered photons is greatly reduced.

\begin{figure}[tb]
\centering
\includegraphics[width=8.0cm]{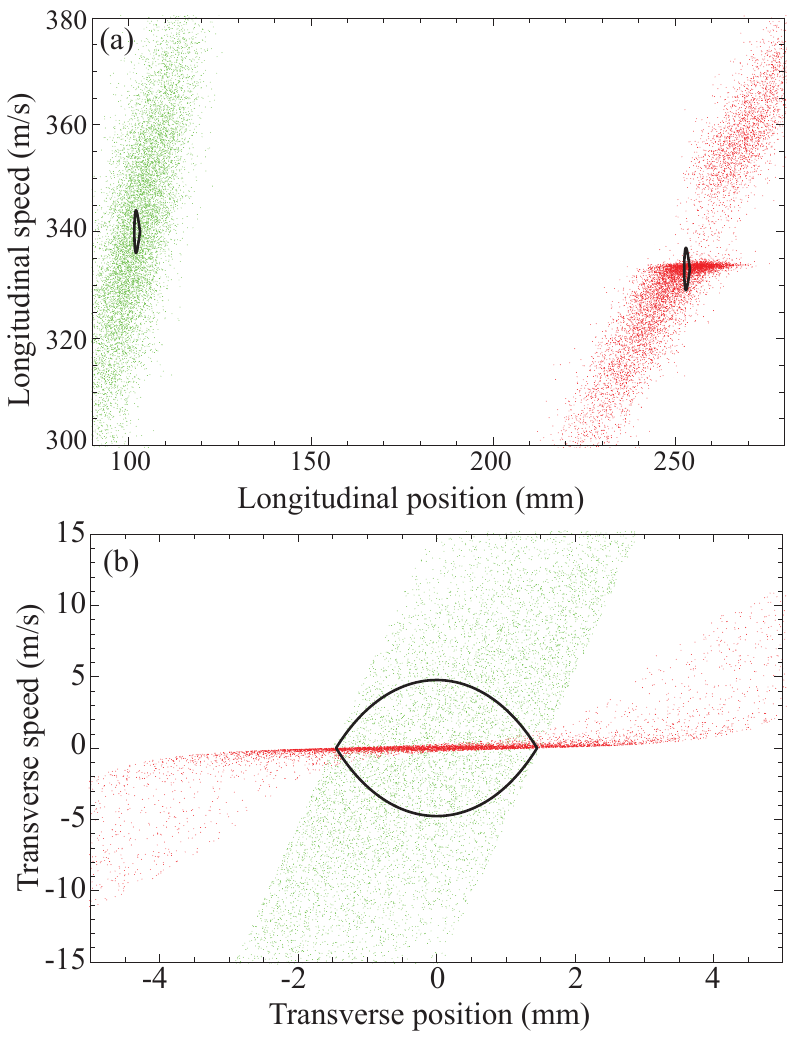}
\caption{Simulations of laser cooling CaF prior to Stark deceleration in a travelling-trap decelerator. The points show the phase-space distributions in (a) the longitudinal direction and (b) the transverse direction. Green points: molecules entering the decelerator at $z=103$\,mm with no laser cooling applied. Red points: molecules entering the decelerator at $z=170$\,mm with laser cooling applied. Line: Phase-space acceptance of the decelerator for CaF($N=1$) decelerated at a rate of $2\times 10^{4}$\,m/s$^{2}$.}
\label{LaserCooling}
\end{figure}

To estimate how effective laser cooling would be in increasing the number of decelerated molecules, we simulate a possible experiment. In this simulation, the laser light consists of two main components, one at 606\,nm resonant with the P(1) line of $\text{A}(v'=0)-\text{X}(v''=0)$, and the other at 628\,nm resonant with the P(1) line of $\text{A}(v'=0)-\text{X}(v''=1)$. Sidebands are applied to address the 4 hyperfine components of each of these transitions, so that there are 8 discrete frequencies in total. A small magnetic field is used to prevent optical pumping into a dark state. The only remaining dark states are the vibrational states with $v''>1$, and the branching ratio for the $v'=0$ state to decay into these dark states is assumed to be approximately $10^{-4}$. Neglecting coherence, the internal dynamics of the molecule interacting with this set of laser beams is calculated using a rate model that includes all the Zeeman sub-levels (24 in the ground state, 4 in the upper state). For the decay rate and the $0-0$ Franck-Condon factor we use the measured values \cite{Wall(1)08}, $\Gamma/(2\pi)=8.3$\,MHz, and $Z=0.987$. The branching ratios between all the sub-levels are calculated using the angular momentum algebra written down in \cite{Wall(1)08}. To include the influence of the magnetic field, which does not fit comfortably in a rate model, we add terms to the rate equations that damp out differences in population between Zeeman sub-levels with a damping rate given by the relevant Larmor precession rate. From these calculations we find that when the intensity at each of the 8 frequencies is $I$, and the frequencies have a common detuning, the photon scattering rate is like that of a two-level system with an effective decay rate $\Gamma_{\text{eff}}/(2\pi)=2.4$\,MHz and an effective saturation parameter $s_{\text{eff}}=I/I_{\text{s,eff}}$, where $I_{\text{s,eff}}\simeq 50$\,mW/cm$^{2}$. We use this effective scattering rate to calculate the scattering forces on the molecule.

We take a supersonic beam of CaF with a mean speed of 340\,m/s and a translational temperature of 3\,K. The spatial distribution of the source is Gaussian in all directions with a standard deviation of 5\,mm in the longitudinal direction and 2.5\,mm in the transverse direction. The beam passes through a 3\,mm diameter skimmer at $z=68$\,mm and then into the decelerator. We consider using the travelling-trap decelerator demonstrated in Berlin \cite{Osterwalder(1)10} to decelerate CaF molecules in the $N=1$ state, these being the ones that are subjected to laser cooling. In the travelling-trap decelerator the molecules are confined in a three-dimensional potential well throughout the deceleration process and the instabilities that arise in the WF decelerator \cite{Sawyer(1)08} are absent. We take the geometry to be exactly that of \cite{Osterwalder(1)10} and the peak-to-peak amplitude of the applied waveform to be 16\,kV. In the case where there is no laser cooling applied we place the decelerator at $z=103$\,mm, close to the exit of the skimmer. In the laser cooling case the entrance of the decelerator is moved downstream to $z=253$\,mm. Cooling is applied in both transverse directions in the 150\,mm gap between the exit of the skimmer and the entrance of the decelerator using retro-reflected laser beams with a common red-detuning of 7.5\,MHz and with $s_{\text{eff}}=0.5$. Longitudinal cooling is applied using a laser beam that propagates along $-z$, has $s_{\text{eff}}=0.5$, and is pulsed on for 550\,$\mu$s starting 200\,$\mu$s after the molecules are produced, so that they interact with the laser in the region between the skimmer entrance and the decelerator entrance. When this light turns on, the common detuning is -578\,MHz, and then the frequency is chirped linearly at a rate of 0.048\,MHz/$\mu$s so that the total frequency change is 26\,MHz. The decelerator traps have an acceleration of $-2\times 10^{4}$\,m/s$^{2}$. Their initial speed is chosen as 340\,m/s for the no-cooling case and $333$\,m/s for the cooling case.

The results of these simulations are shown in Fig.\,\ref{LaserCooling}, where we plot the phase-space distributions of molecules entering the decelerator in the two cases, and compare these to the phase-space acceptance of the decelerator in both the longitudinal and transverse directions. In the longitudinal direction the laser slows down molecules with speeds in the 335--345\,m/s range, substantially compressing the speed distribution into a narrow bunch centred near 335\,m/s. The laser cooling does nothing to compress the spatial distribution which continues to expand so that at the decelerator entrance the laser-cooled bunch is longer than the uncooled bunch. The over-all effect of the longitudinal laser cooling is to increase the number of molecules within the decelerator acceptance by a factor of 3.1. The picture is similar for transverse cooling: the cooling is very effective at reducing the transverse velocity spread, but the spatial spread increases because of the extra distance the molecules have to travel. The number of molecules overlapping with the transverse acceptance increases by a factor of 1.4. With the laser cooling applied in all three dimensions the number of decelerated molecules is predicted to increase by a factor of 6. The decelerated molecules occupy a far smaller volume of phase space when the cooling is applied, because in the transverse direction the velocity spread is compressed to an area far smaller than the acceptance area. The simulations show that the phase-space density of the molecules exiting the decelerator increases by a factor of about 2000 when the laser cooling is applied. So we see that although the laser cooling results in only a modest increase in the number of decelerated molecules, it greatly increases the phase-space density of slow molecules. A further increase in phase-space density could be obtained by using a second short pulse of longitudinal cooling as the molecules exit the decelerator.

\section{Conclusions}

We have reported the Stark deceleration of CaF molecules for the first time and have compared the performance of two types of decelerator for slowing these molecules. The AG decelerator is used to slow down molecules in the strong-field seeking ground state, whereas the WF decelerator works only for molecules in weak-field seeking states and has been applied here to CaF in the $(N,M)=(4,0)$ state.

Using a 100-stage WF decelerator at 20\,kV and at a phase angle of $75^{\circ}$, the molecules have been decelerated from 340\,m/s to 224\,m/s. A 200-stage decelerator of the same design would be able to bring these molecules to rest. For this case, the relatively high phase angle results in a small acceptance. This could be increased by using a longer decelerator at a lower phase angle. A travelling-trap decelerator of the type recently demonstrated in Berlin \cite{Osterwalder(1)10} is also suitable for decelerating CaF and should increase the number of slow molecules obtained. The WF or travelling trap decelerator is also suitable for slowing down molecules of considerably higher mass than CaF, although as the mass increases higher-lying rotationally-excited states are needed for efficient deceleration, and the required decelerator length increases. The required length could be reduced by the use of cryogenically-cooled sources \cite{Maxwell(1)05} to provide slower, more intense beams. Cryogenic sources to date produce either continuous beams or very long pulses that are not well matched to the small size of the potential wells in a typical decelerator. However, there does not seem to be any barrier to making far shorter pulses.

Using a 21-stage AG decelerator at 20\,kV, ground state molecules have been decelerated from 433\,m/s to 399\,m/s. We observe a smaller decelerated signal than expected from numerical simulations, and in the case where all 21 stages are used the decelerated bunch is moving a little faster than expected. The advantage of the AG decelerator is its ability to focus molecules in strong-field seeking states. For large molecules, these are the only states available \cite{Wohlfart(1)08}. There are several difficulties in decelerating large molecules to low velocity. Firstly, the molecules must stay inside the stability region for AG focussing and that requires either the lens length or the focussing power to change as the molecules slow down. One approach is to use long lenses at the start of the decelerator and short ones at the end, with the long lenses segmented into a series of shorter ones so as to maintain the same deceleration per unit length. This arrangement has been discussed in detail in the context of decelerating ground state YbF \cite{Tarbutt(1)09}. An alternative is to build each lens from four electrodes instead of two \cite{Bethlem(1)06} in which case the focussing (and defocussing) direction of each lens is set by the polarities of the electrodes. This provides a more versatile structure because the exact arrangement of lenses is no longer fixed by the layout of the machine. Even with these more sophisticated lens arrangements it becomes very difficult to maintain stability of AG focussing at the lowest speeds \cite{Tarbutt(1)09}. A further difficulty for strong-field seeking molecules is that AG deceleration is always accompanied by excess defocussing, as discussed above. This can be minimized by tapering the ends of the electrodes more gradually, for example by replacing the hemispherical ends of the electrodes in the present design with ellipsoidal ends. Finally, experimental work on AG deceleration and focussing consistently reports more loss than expected from detailed numerical simulations, and these additional losses are thought to be due to mechanical misalignments. Exceptionally good alignment is needed to make the decelerator perform as expected. A cryogenically-cooled source operated in the effusive regime offers an alternative, and simpler, method to obtain slow-moving beams of large, heavy molecules.

For some molecules, CaF being a good example, laser cooling promises to increase the phase-space density enormously. Laser cooling can both cool and decelerate the molecules. For the latter, a large number of photons have to be scattered, requiring a particularly `closed' vibrational structure. An alternative, requiring far fewer spontaneous emissions, is to combine laser cooling with Stark deceleration. We have simulated a possible experiment where transverse and longitudinal laser cooling is applied to CaF prior to Stark deceleration. The cooling is very effective in compressing the velocity distribution, but does not compress the spatial distribution. Instead, the size of the beam at the decelerator entrance increases because the decelerator needs to be moved downstream to allow space for laser cooling. In our simulations, the laser cooling increases the number of decelerated molecules by a factor of about 6, but increases the phase-space density of slow molecules by about 2000.

\acknowledgments We are grateful to the EPSRC, the STFC and the Royal Society for supporting this work.

\end{document}